\documentclass[prb,showpacs,amsmath,amssymb,twocolumn]{revtex4}

\usepackage{graphicx}
\usepackage{dcolumn}
\usepackage{bm}
\begin{document}

\title{Doping and disorder in the Co$_2$MnAl and Co$_2$MnGa half-metallic Heusler alloys}

\author{K. \"Ozdo\~gan}\email{kozdogan@gyte.edu.tr}
\affiliation{Department of Physics, Gebze Institute of Technology,
Gebze, 41400, Kocaeli, Turkey}

\author{E. \c Sa\c s\i o\~glu}\email{e.sasioglu@fz-juelich.de}
\affiliation{Institut f\"ur Festk\"orperforschung,
Forschungszentrum
J\"ulich, D-52425 J\"ulich, Germany\\
Fatih University,  Physics Department, 34500,    B\" uy\" uk\c
cekmece,  \.{I}stanbul, Turkey}

\author{B. Akta\c s}
\affiliation{Department of Physics, Gebze Institute of Technology,
Gebze, 41400, Kocaeli, Turkey}

\author{I. Galanakis}\email{galanakis@upatras.gr}
\affiliation{Department of Materials Science, School of Natural
Sciences, University of Patras,  GR-26504 Patra, Greece}

\date{\today}

\begin{abstract}
We expand our study on the full-Heusler compounds [I. Galanakis
\textit{et al.}, Appl. Phys. Lett. \textbf{89}, 042502 (2006)] to cover also the
case of doping and disorder in the case of Co$_2$MnAl and
Co$_2$MnGa half-metallic Heusler alloys. These alloys present a
region of very small minority density of states instead of a real
gap. Electronic structure calculations reveal that doping with Fe
and Cr in the case of Co$_2$MnAl retains the half-metallicity
contrary to the Co$_2$MnGa compound. Cr impurities present an
unusual behavior and the spin moment of the Cr impurity scales
almost linearly with the concentration of Cr atoms contrary to the
Co$_2$MnZ (Z= Si, Ge, Sn) where it was almost constant.
Half-metallicity is no more preserved for both Co$_2$MnAl and
Co$_2$MnGa alloys when disorder occurs and there is either excess
of Mn or $sp$ atoms.
\end{abstract}


\maketitle

Research on half-metallic ferromagnets is rapidly growing since
its prediction for NiMnSb in 1983 by de Groot and his
collaborators.\cite{deGroot} The driving force of the interest in
these materials is their potential applications in
magnetoelectronic
applications.\cite{book,Westerholt,Reiss,Sakuraba,Dong} Electronic
structure calculations have played a central role on the study of
these materials. Several new half-metallic ferromagnetic materials
and their properties have been initially predicted by theoretical
ab-initio calculations and later verified by experiments. Among the
half-metallic materials, intermetallic Heusler alloys have
attracted considerable attention due their easy growth and the
high Curie temperatures.\cite{Galanakis,Sasioglu}

The full-Heusler compounds containing Co and Mn are of particular
interest for spintronics since they combine high Curie
temperatures and coherent growth on top of semiconductors (they
consist of four f.c.c. sublattices with each one occupied by a
single chemical element). To precisely control the
properties of these compounds we have to study effects susceptible
of inducing states within the minority spin gap and thus destroy
the half-metallicity. States at the interfaces of these compounds
with semiconductors\cite{Interfaces} as well as temperature driven
excitations\cite{Chioncel,MavropTemp,Skomski} seem to destroy
half-metallicity. In addition to  interface states and
temperature, the third main effect which can destroy
half-metallicity is the appearance of defects and disorder.

In a recent paper\cite{APL} we studied the effect of doping and
disorder on the magnetic properties of the Co$_2$MnSi, Co$_2$MnGe,
Co$_2$MnSn full-Heusler alloys. Doping simulated by the
substitution of Cr and Fe for Mn in these alloys overall keeps the
half-metallicity. The effect of doping depended clearly on the
position of the Fermi level, having the largest one in the case of
Co$_2$MnSi where the Fermi level is near the edge of the
minority-spin gap. On the other hand disorder, simulated either by
excess of Mn atoms at the D site occupied in the perfect compound
by the  $sp$ atoms or \textit{vice versa}, was found to be more
important for the heavy $sp$ atoms like Sn but in general the
half-metallicty was almost preserved.

In this manuscript we expand our theoretical work to include also
the case of Co$_2$MnAl and Co$_2$MnGa compounds which have one
valence electron less than that of the previous ones. Doping by
electrons simulated by partial substitution of Fe for Mn in the
unit cell has little effect on the gap-properties while Cr
impurities, corresponding to hole doping, exhibit a more unusual
behavior. A high degree of spin-polarization at the Fermi level
is overall preserved but contrary to what happens for the Si, Ge
and Sn compounds and to the Fe-doping case, Cr impurities present
a very small spin moment for small concentrations which increases
with the concentration.
 In the last part of our
study we concentrate on the case of disorder which is shown to
severely affect the half-metallicity contrary to the Co$_2$MnZ
(Z=Si, Ge, Sn) alloys. The electronic structure calculations are
performed using the full--potential nonorthogonal local--orbital
minimum--basis band structure scheme (FPLO).\cite{koepernik}

\begin{figure}
\includegraphics[scale=0.5]{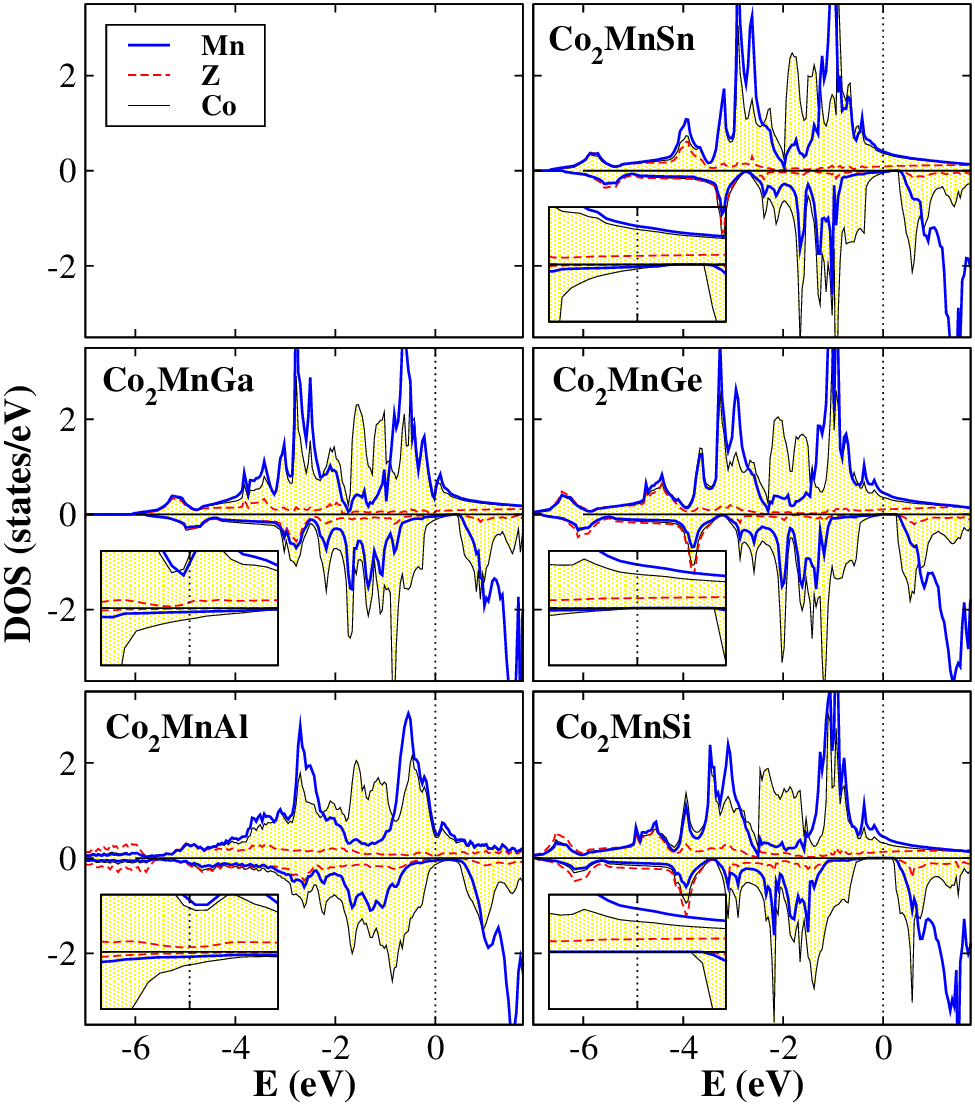}
\caption{(Color online) Atom-resolved density of states (DOS) for
the Co$_2$MnZ compounds, where Z is Al, Ga, Si, Ge, and Sn.  We
have set the Fermi level as the zero of the Energy axis. In the
onsets we have blown up the region around the Fermi level. Note
that positive values of DOS refer to the majority-spin electrons
and negative values to the minority-spin electrons. \label{fig1}}
\end{figure}

Before presenting our results we have drawn in Fig. \ref{fig1} the
atom-resolved DOS for the Co$_2$MnZ compounds. On the left column
are the case with Z a $sp$ element belonging to the IIIB column of
the periodic table (Al and Ga) and in the right column the case of
IVB elements (Si, Ge and Sn). The extra electron in the latter
case occupies majority states leading to an increase of the
exchange splitting between the occupied majority and the
unoccupied minority states and thus to larger gap-width for the
Si-, Ge- and Sn-based compounds. In the case of Al- and Ga-based
alloys the bonding and antibonding minority d-hybrids almost
overlap and the gap is substituted by a region of very small
minority density of states (DOS); we will call it a pseudogap. In
all cases the Fermi level falls within the gap or the pseudogap
and an almost perfect spin-polarization at the Fermi level is
preserved.

\begin{figure}
\includegraphics[scale=0.53]{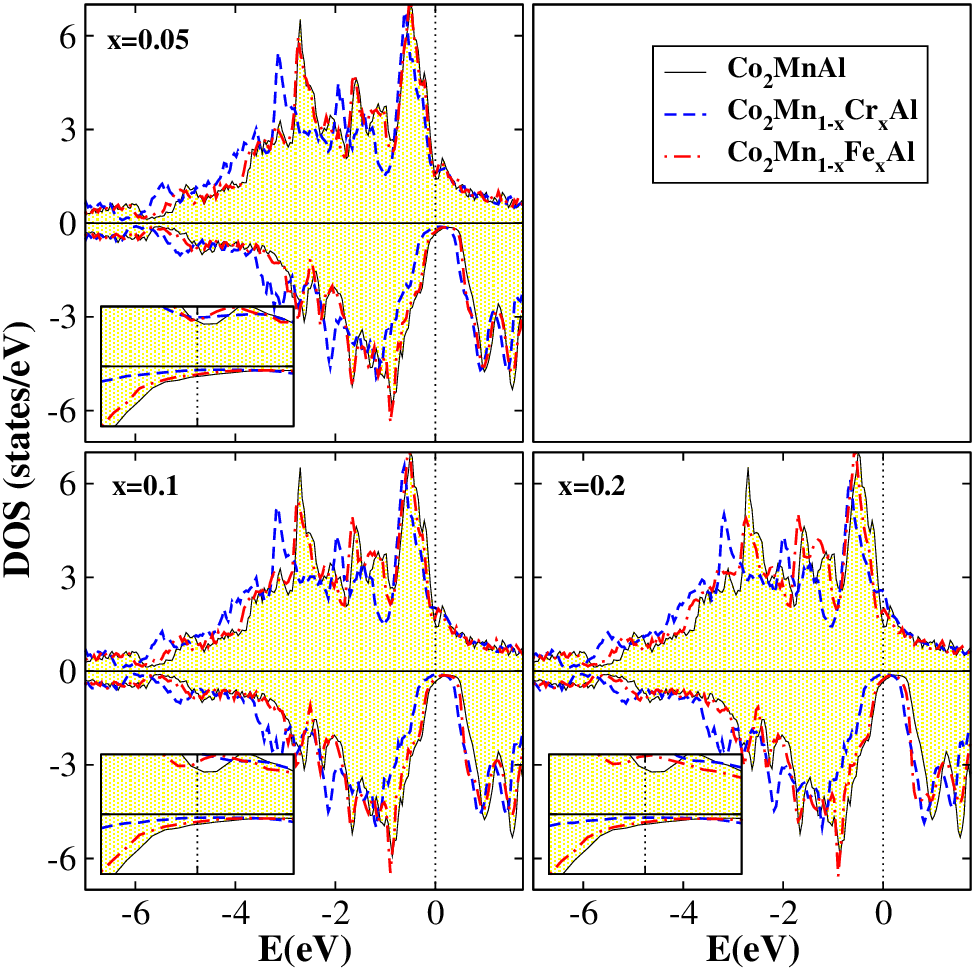}
\caption{(Color online) Spin-resolved DOS for the case of
Co$_2$Mn$_{1-x}$Cr$_x$Al and Co$_2$Mn$_{1-x}$Fe$_x$Al for three
values of the doping concentration $x$. DOS's are
compared to the one of the undoped Co$_2$MnAl alloy. Details as in
Fig. \ref{fig1}. \label{fig2}}
\end{figure}

Firstly we will study the doping in these compounds. We substitute
either Fe or Cr for Mn to simulate the doping by electrons and
holes, respectively.  We study the cases of moderate doping by
substituting 5\%, 10\% and 20\% of the Mn atoms. In our
calculations the use of coherent potential approximation  ensures
doping in a random way. In Fig. \ref{fig2} we present the total
density of states (DOS) for the Co$_2$Mn$_{1-x}$(Fe or Cr)$_x$Al
alloys to compare to the perfect  Co$_2$MnAl alloys and in Table
\ref{table1} we have gathered the total and atom-resolved spin
moments for both Al- and Ga-based alloys. Note that in the figure
we have blown up in the onsets the region around the Fermi level
where the gap exists.

\begin{table*}
\caption{Total and atom-resolved spin magnetic moments for the
case of Fe and Cr doping of the Mn site in $\mu_B$. The total
moment in the cell is the sum of the atomic ones multiplied by the
concentration of this chemical element. Note that for Cr, Mn and Fe we have scaled the spin moments to one atom
and that for Co we
give the sum of the moments of both atoms.} \label{table1}
\begin{ruledtabular}
 \begin{tabular}{l|ccccc|ccccc||l|ccccc|ccccc}
    &    Total  &   Co &   Mn&Cr&
 sp &  Total &
Co &   Mn&Fe&  sp &    &    Total  &   Co &   Mn&Cr&
 sp &  Total &
Co &   Mn&Fe&  sp  \\ \hline $x$&
\multicolumn{5}{c|}{Co$_2$Mn$_{1-x}$Cr$_{x}$Al} &
\multicolumn{5}{c||}{Co$_2$Mn$_{1-x}$Fe$_{x}$Al}
&$x$& \multicolumn{5}{c|}{Co$_2$Mn$_{1-x}$Cr$_{x}$Si} & \multicolumn{5}{c}{Co$_2$Mn$_{1-x}$Fe$_{x}$Si} \\
  0.00  &  4.04& 1.36& 2.82&      & -0.14& 4.04& 1.36& 2.82&      & -0.14 & 0.00 &   5.00 &   1.96 &    3.13 &
    &      -0.09 &   5.00  & 1.96&   3.13   &       & -0.09\\
  0.05  &  3.95& 1.49& 2.69& 0.34 & -0.13& 4.06& 1.44& 2.76& 2.78 &
  -0.13 &  0.05  &  4.95 &   1.97  &   3.12 &    2.06 &    -0.09 &   5.05   & 2.02   &  3.13    &   2.87  &
  -0.09\\
  0.10  &  3.90& 1.51& 2.71& 0.62 & -0.11& 4.11& 1.49& 2.76& 2.78 &
  -0.13 &  0.10  &  4.90&   1.97   &  3.12  &   2.07   &  -0.09&   5.09   & 2.06    & 3.17  &   2.85  &
  -0.08\\
  0.20  &  3.80& 1.54& 2.74& 0.91 & -0.11& 4.21& 1.58& 2.76& 2.79 &
  -0.13 & 0.20  &  4.80&    1.97  &   3.12 &    2.09    &  -0.08&   5.14   & 2.13 &  3.16  &   2.82   &
  -0.08\\
 $x$&\multicolumn{5}{c|}{Co$_2$Mn$_{1-x}$Cr$_{x}$Ga} &\multicolumn{5}{c||}{Co$_2$Mn$_{1-x}$Fe$_{x}$Ga}  &
  $x$&\multicolumn{5}{c|}{Co$_2$Mn$_{1-x}$Cr$_{x}$Ge} &\multicolumn{5}{c}{Co$_2$Mn$_{1-x}$Fe$_{x}$Ge} \\
 0.00 & 4.09 &  1.30&  2.88&       &  -0.10& 4.09 &  1.30&  2.88&       &
 -0.10 & 0.00   & 5.00   &1.87  & 3.20    &         & -0.06 &  5.00 &  1.87  &  3.20   &        &    -0.06\\
 0.05 & 4.05 &  1.36&  2.93&   0.20&  -0.10& 4.15&  1.36& 2.90&   2.76&
 -0.10 & 0.05   & 4.95  & 1.86  & 3.21 & 2.05   & -0.06&  5.05  &  1.91  &  3.22 &   2.88 &
  -0.06\\
 0.10 & 4.00 &  1.39&  2.94&   0.55&  -0.10& 4.20&  1.41& 2.90&   2.76&
 -0.10 & 0.10   & 4.90 &  1.86  & 3.22    & 2.07   & -0.06&  5.10 &  1.96  &  3.23  &   2.88  &   -0.06\\
 0.20 & 3.88 &  1.42&  2.97&   0.91&  -0.09& 4.30&  1.52& 2.92&   2.77&
 -0.10& 0.20    &4.80&   1.86  & 3.22   & 2.10   & -0.06&  5.19  &  2.06  &  3.26  &  2.89   &   -0.05
\end{tabular}
\end{ruledtabular}
\end{table*}

\begin{table}
\caption{Total and atom-resolved spin magnetic moments for the
case of Cr doping of the Mn site for both Al- and Si-based compounds in $\mu_B$. Details as in
Table \ref{table1}.} \label{table2}
\begin{ruledtabular}
 \begin{tabular}{l|ccccc|ccccc}
$x$& \multicolumn{5}{c|}{Co$_2$Mn$_{1-x}$Cr$_{x}$Al} & \multicolumn{5}{c}{Co$_2$Mn$_{1-x}$Cr$_{x}$Si}  \\
   &    Total  &   Co &   Mn&Cr&
 sp &  Total & Co &   Mn&Cr&  sp \\ \hline
  0.00  &  4.04& 1.36& 2.82&        &  -0.14&5.00&   1.96&    3.13  &     &
  -0.09\\
  0.10  &  3.90& 1.51& 2.71& 0.62   &  -0.11&4.90&   1.97&   3.12   &   2.07    &
  -0.09\\
  0.20  &  3.80& 1.54& 2.74& 0.91   &  -0.11&4.80&   1.97&     3.12 &    2.09    &
  -0.08\\
  0.30  &  3.70&   1.55& 2.76& 1.09 &  -0.10&4.70&   1.96&     3.12 &    2.11    &
  -0.08\\
  0.40  &  3.60&   1.55& 2.77& 1.20 &  -0.10&4.60&   1.95&     3.12&    2.12    &
  -0.08\\
  0.50  &  3.50&   1.55& 2.78& 1.29 &  -0.09&4.50&   1.94&     3.13&    2.14    &
  -0.07\\
  0.60  &  3.40&   1.54& 2.79& 1.37 &  -0.09&4.40&   1.93&     3.13 &    2.15    &
  -0.07\\
  0.70  &  3.30&   1.54& 2.80& 1.43 &  -0.08&4.30&   1.92&     3.13 &    2.16    &
  -0.07\\
  0.80  &  3.20&   1.53& 2.83& 1.48 &  -0.08&4.20&   1.91&     3.13  &    2.17    &
  -0.07\\
  0.90  &  3.10&   1.51& 2.84& 1.53 &  -0.07&4.10&   1.90&    3.12  &   2.17     &
  -0.07\\
  1.00  &  3.00&   1.46&     & 1.63 &  -0.09&4.00&   1.89&                &  2.17       & -0.06
\end{tabular}
\end{ruledtabular}
\end{table}

Fig. \ref{fig2} confirms the  discussion in Ref.
\onlinecite{Galanakis} that the gap is created between states
located exclusively at the Co sites. As was the case also for the
compounds in Ref. \onlinecite{APL} the majority-spin occupied
states form a common Mn-Co band while the occupied minority states
are mainly located at the Co sites and minority unoccupied at the
Mn sites. Doping the perfect ordered alloy with Fe has only a
marginal effect on the total DOS compared to the more significant
effects of the doping with Cr.  In the latter case it seems that
Cr-doping slightly opens the pseudogap and pushes both the
majority and minority occupied bands lower in energy. This
behavior upon Cr doping is also reflected on the spin moment of
the Cr impurity atoms as we will discuss latter in the manuscript
and is in contrast to what happens upon Cr-doping for the
Co$_2$MnSi compound presented in Ref. \onlinecite{APL} where the
DOS scarcely changed. The important detail is what happens around
the Fermi level and in what extent is the gap in the minority band
affected by the doping. So now we will concentrate only at the
enlarged regions around the Fermi level. The blue dashed lines
represent the Cr-doping while the red dash-dotted lines are the
Fe-doped alloys. The situation is reversed with respect to the
Co$_2$MnSi compound, Cr-doping has significant effects to the
pseudogap. Its width is larger with respect to the perfect
compound and becomes slightly narrower as the degree of the doping
increases. We will discuss this behavior in detail later in the
text.  Fe-doping on the other hand almost does not change the DOS
around the Fermi level. The extra-electrons occupy high-energy
lying antibonding majority states but since Co$_2$MnAl has one
valence electron less than Co$_2$MnSi half-metallicity remains
energetically favorable and no important changes occur upon
Fe-doping and   further substitution of Fe for Mn should retain
the half-metallicity even for the Co$_2$FeAl compound contrary to
the Co$_2$FeSi compound which is not half-metallic.\cite{GalaQuat}

\begin{figure}
\includegraphics[scale=0.5]{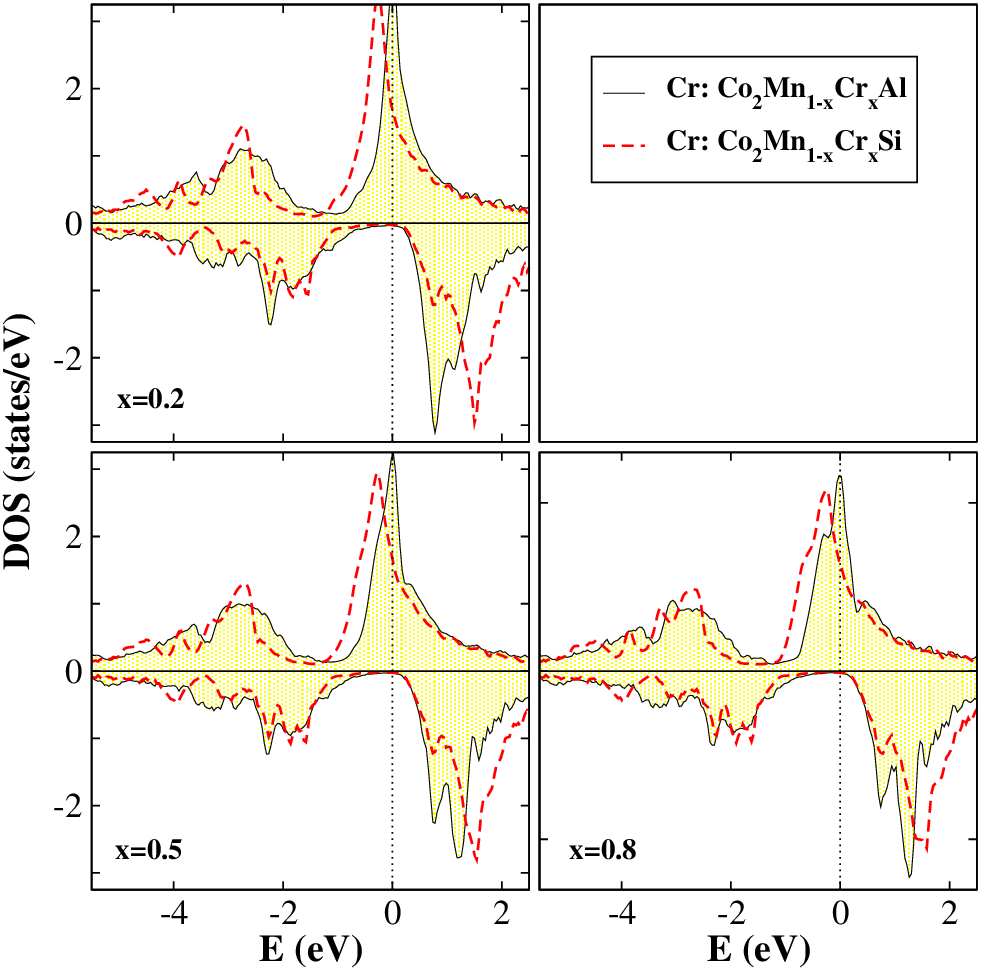}
\caption{(Color online) Cr-resolved DOS in the case of the
Co$_3$Mn$_{1-x}$Cr$_x$Al and Co$_3$Mn$_{1-x}$Cr$_x$Si compounds
for three different values of the concentration $x$. Details as in
Fig. \ref{fig1}. \label{fig3}}
\end{figure}

In Table \ref{table1} we have gathered the spin magnetic moments
for all cases under study. For perfect half-metals the total spin
moment $M_t$  follows the Slater-Pauling (SP) behavior being the
number of the valence electrons in the unit cell minus
24.\cite{Galanakis} The perfect Co$_2$MnAl and Co$_2$MnGa
compounds have total spin moments slightly larger than the ideal 4
$\mu_B$ predicted by the SP rule while the total spin moments of the
Si- and Ge-based alloys are exactly 5 $\mu_B$ as predicted by the SP rule.
In the case of the chemically
disordered compounds, doping by 5\%, 10\% or 20\% of Cr (or Fe)
atoms, means that the mean value of the total number of valence
electrons in the unit cell is decreased (or increased
respectively) by 0.05, 0.10 and 0.20 electrons, respectively. The
half-metallicity retains only when we dope Co$_2$MnAl.  On the
other hand for the Co$_2$MnGa compound the Fermi level is deeper
in energy as shown in Fig. \ref{fig1} and both perfect and
disordered compounds are not half-metallic.

The atom-resolved spin moments in Table \ref{table1} present a
striking peculiarity when we dope with Cr. In the case of
Co$_2$MnSi(or Ge) compounds,  the Cr
impurities have a moment of slightly larger than 2 $\mu_B$
independent of the degree of doping. Similarly Fe impurities have a
spin moment of around 2.8 $\mu_B$. In the case of the Co$_2$MnAl
and Co$_2$MnGa compounds Fe impurities have also a large spin
moment of the same order of magnitude. But Cr  impurities have a
very small spin moment (0.3 $\mu_B$ for the Al compounds and 0.2
$\mu_B$ for the Ga compound) for the case of 5\%\ doping.
Substituting 10\%\ or 20\%\ of Cr for the the Mn atoms leads to a
doubling or tripling, respectively, of the Cr-impurity spin
moment. To elucidate this behavior, we have performed calculations
for both Co$_2$[Mn$_{1-x}$Cr$_x$]Al and Co$_2$[Mn$_{1-x}$Cr$_x$]Si
compounds for values of $x$ ranging from 0 to 1 and we have
gathered the atom-resolved spin moment in Table \ref{table2} and
in Fig. \ref{fig3} we have drawn the Cr-resolved DOS for three
values of $x$. In the case of Si-compounds Cr-impurities spin
moment is almost constant irrespectively of the concentration. On
the other hand for the Al-compounds the spin moment of the
Cr-impurities increases with the concentration of the Cr atoms.
Fig. \ref{fig3} explains this phenomenon. In the case of the
Si-compounds the Fermi level falls after the majority peak  and
upon doping the relative position of the Fermi level does not
change. This peak comes from the $d$-electrons of the Cr
containing also the double degenerated $e_g$ and triple
degenerated $t_{2g}$ $d$-electrons. In the case of the
Al-compounds the Fermi level is pinned exactly at the peak of the
$e_g$ electrons which are more localized and as we dope the
compounds we change slightly the Coulomb repulsion and the
exchange splitting and the triple degenerated $t_{2g}$ electrons,
which spread over a wider energy  region, move lower in energy
creating also the shoulder presented for $x=0.8$.

In the last part of our study we study the effect of disorder. We
either create an excess of the Mn or the  $sp$ atoms. In Table
\ref{table3} we have gathered the total and atomic spin moments
for all cases under study. Substituting 5\%, 10\%, 15\% or 20\% of
the Mn atoms by the Al or Ga  ones, corresponding to the negative
values of $x$ in the table, results in a decrease of 0.15, 0.30,
0.45 and 0.60 of the total number of valence electrons in the
cell, while the inverse procedure results to a similar increase of
the mean value of the number of valence electrons. Contrary to the
Si, Ge and Sn compounds presented in Ref. \onlinecite{APL} which
retained the perfect half-metallicity, the Al- and Ga-based
compounds are no more half-metallic. In the case of the Si and
related compounds disorder induced states at the edges of the gap
keeping the half-metallic character but this is no more the case
for the Al and Ga compounds where no real gap exists.

\begin{table}
\caption{Total and atom-resolved spin magnetic moments for the
case of excess of Mn ($x$ positive) or $sp$ atoms ($x$ negative)
atoms. In the second column the ideal total spin moment if the
compound was half-metallic. Details as in Table \ref{table1}.}
\label{table3}
\begin{ruledtabular}
 \begin{tabular}{r|c|cccc|cccc}
 & & \multicolumn{4}{c|}{Co$_2$Mn$_{1+x}$Al$_{1-x}$}
 & \multicolumn{4}{c}{Co$_2$Mn$_{1+x}$Ga$_{1-x}$} \\ \hline
  $x$ & Ideal &   Total  &   Co &   Mn&  Al &    Total  &   Co &   Mn&  Ga  \\ \hline
 -0.20  &  3.40 & 3.26 &   1.09&   2.89  &  -0.12& 3.40 &  1.11 &  3.00 &
 -0.09\\
 -0.15  &  3.55 & 3.45 &   1.15&   2.87  &  -0.12& 3.57 &  1.15 &  2.98 &
 -0.10\\
 -0.10  &  3.70 & 3.64 &   1.22&   2.84  &  -0.13& 3.74 &  1.21 &  2.94 &
 -0.10\\
 -0.05  &  3.85 & 3.83 &   1.29&   2.83  &  -0.13& 3.92 &  1.25 &  2.93
& -0.10\\
  0.00  &  4.00 & 4.04 &   1.36&   2.82  &  -0.14& 4.09 &  1.31 &  2.88 &
  -0.10\\
  0.05  &  4.15 & 4.22 &   1.40&   2.81  &  -0.14& 4.29 &  1.36 &  2.89 &
  -0.11\\
  0.10  &  4.30 & 4.40 &   1.44&   2.81  &  -0.14& 4.48 &  1.40 &  2.88 &
  -0.11\\
  0.15  &  4.45 & 4.60 &   1.49&   2.81  &  -0.14& 4.66 &  1.45 &  2.88 &
  -0.11\\
  0.20  &  4.60 & 4.80 &   1.54&   2.81  &  -0.15& 4.85 &  1.50 &  2.87 &
  -0.11
\end{tabular}
\end{ruledtabular}
\end{table}

We have studied the effect of doping and disorder on the magnetic
properties of the Co$_2$MnAl and Co$_2$MnGa full-Heusler alloys.
These compounds present a region of low minority density of states
instead of a real gap. Doping simulated by the substitution of Cr
and Fe for Mn overall keeps the half-metallicity for Co$_2$MnAl
while the Ga-compounds present deviations. The spin-moment of the
Cr impurities varies considerably with the concentration and this
behavior is attributed to the position of the majority $t_{2g}$
$d$-electrons with respect to the majority $e_g$ electrons.
Disorder simulated by excess of either the Mn or $sp$ atoms
completely destroys the almost perfect spin-polarization of the
perfect compounds. It seems that Co$_2$MnAl and Co$_2$MnGa
compounds are less adequate than the Co$_2$MnSi, Co$_2$MnGe and
Co$_2$MnSn alloys for realistic spintronic applications.


\end{document}